%% file: main.tex
\documentclass[conference]{IEEEtran}
\IEEEoverridecommandlockouts

\usepackage{silence}
\usepackage[left=1.58cm,right=1.58cm,top=1.91cm,bottom=2.54cm]{geometry}

\usepackage{cite}

\usepackage{amsmath,amssymb,amsthm,fixmath}

\usepackage[T1]{fontenc}
\usepackage{lmodern} 

\usepackage{color,colortbl}
\usepackage{xcolor}

\usepackage[inline]{enumitem}

\usepackage{siunitx}
\DeclareSIUnit{\dBm}{dBm}

\usepackage{graphicx}
\usepackage[labelformat=simple]{subcaption}

\usepackage{epstopdf}
\usepackage{cuted}
\usepackage{multirow,booktabs}
\usepackage{diagbox}
\usepackage{makecell}
\usepackage{hhline}
\usepackage{array} 
\newcolumntype{x}{!{\vrule width 2px}}
\newcolumntype{y}{!{\vrule width 1.5px}}

\usepackage{optidef}

\usepackage[hidelinks]{hyperref}

\usepackage[shortcuts,acronym,automake]{glossaries}
\input{glossary}

\usepackage{tcolorbox}
\tcbuselibrary{many}

\usepackage[norelsize,linesnumbered,ruled]{algorithm2e}
\SetKwRepeat{Do}{do}{while}
\makeatletter
\newcommand{\removelatexerror} {\let\@latex@error\@gobble}
\makeatother

\usepackage[normalem]{ulem}

\usepackage[textsize=tiny,colorinlistoftodos]{todonotes}

\tikzstyle{note}=[rectangle, minimum width=3cm, draw = none, fill = none, minimum width = 1.5cm, anchor=center, align=left]
\tikzstyle{block}=[rectangle, draw, line width=1pt, fill = none, minimum width = 1cm, minimum height = 0.75cm, anchor=center, inner sep = 0.5mm, align=center]
\tikzstyle{arrow} = [thick,->,>=stealth]

\newif\ifreviewmode
\reviewmodetrue 

\ifreviewmode
\else
  \renewcommand{\todo}[1]{} 
\fi

\begin{document}

\title{A Novel Dynamic Epidemic Model for\\Successive Opinion Diffusion in Social Networks}

\author{
	\IEEEauthorblockN{
		Bin~Han\IEEEauthorrefmark{1},
		Fabienne~Renckens\IEEEauthorrefmark{1},
		C.~Clark~Cao\IEEEauthorrefmark{2},
		and~Hans~D.~Schotten\IEEEauthorrefmark{1}\IEEEauthorrefmark{3}
	}
	\IEEEauthorblockA{
		\IEEEauthorrefmark{1}RPTU Kaiserslautern-Landau,
		\IEEEauthorrefmark{2}Lingnan University, 
		\IEEEauthorrefmark{3}German Research Center for Artificial Intelligence (DFKI)
	}
}

\maketitle

\begin{abstract}
This paper proposes a dynamic epidemic model for successive opinion diffusion in social networks, extending the SHIMR model. It incorporates dynamic decision-making influenced by social distances and captures accumulative opinion diffusion caused by interrelated rumors. The model reflects the impact of rumor spread on social network structures. Simulations validate its effectiveness in explaining phenomena like the echo chamber effect and provide insights into opinion diffusion dynamics, with implications for understanding social polarization and network evolution.
\end{abstract}

\begin{IEEEkeywords}
Social network, epidemic model, rumor
\end{IEEEkeywords}

\section{Introduction}
Since the beginning of this century, online social media networks have dramatically developed, substantially changing the paradigm of information dissemination. The rapid spread of information, opinions, and rumors in social networks has attracted increasing attention from researchers in sociology, psychology, and computer science. Studying information diffusion in social networks is crucial for understanding their dynamics and has practical implications for the public.

Existing models of the information diffusion process in social networks can be clustered into two categories: explanatory and predictive. Explanatory models aim to understand the underlying mechanisms of information diffusion in social networks, while predictive models focus on predicting the future spread of information in social networks. More specifically focusing on the former category, most explanatory models can be further classified into two types: the epidemic models that consider the information difussion process similar to the epidemic spread process, and the influence models that consider the information diffusion process as a result of the influence of individuals on each other~\cite{LWG+2017survey}. 

Various conventional explanatory models have been developed and widely applied to capture the dynamic process of individual information spreading in social networks. However, they generally consider the social network as a staic environment and overlook the impact that information can make to the network itself. Nor have they taken into account the correlation among multiple rumors with respect to the same subject or shared ideology. Practically, however, it has been observed that that the spread of rumors, triggering individuals' different reactions to them, can lead to changes in social relations among these individuals, affect the structure of the social network as well as the spread of succeeding information, and eventually cause accumulated diffusion of opinion. One typical example of such phenomena is the well-known ``echo chamber'' effect~\cite{CDG+2021echo}, where individuals with similar opinions tend to form a closed group and reinforce each other's opinions, leading to the polarization and division of the society. Conventional explanatory models, unfortunately, fail to encompass the advanced dynamic processes of such kind.

In this paper, we prpose a novel epidemic model that extends the conventional SHIMR model~\cite{LWO2022rumor} by considering a dynamic decision scheme of each individual, which is both influencing and influenced by its social distances to the other individuals in the network. The proposed model is able to capture the accumulative opinion diffusion caused by multiple successive rumors sharing a same subject or ideology, and to reflect the impact of the spread of rumors on the social network structure. We conduct numerical simulations to validate the effectiveness of the proposed model, and discuss the implications of the model in understanding the underlying mechanism and dynamics of opinion diffusion in social networks.

The remainder of this paper is organized as follows: first we review several conventional epidemic models in Section~\ref{sec:sota}, then in Section~\ref{sec:model} we introduce our proposed dynamic SHIMR model. Subsequently, in Section~\ref{sec:simualtion} we introduce our numerical simulations and the results, together with discussions. In the end, we conclude the paper with Section~\ref{sec:conclusion}.

\section{Conventional Epidemic Models}\label{sec:sota}
Research interest in modeling the epidemic spread process dates back to the early 20\textsuperscript{th} century, when \emph{Kermack} and \emph{McKendrick} proposed the first mathematical model to describe the spread of infectious diseases~\cite{KM1927contribution}. This famous SIR model divides the population into three compartments: susceptible (S), infected (I), and removed (R). Each individual can transit from the susceptible state to the infected state with a certain infection rate, and from the infected state to the removed state with a certain recovery rate. Once entered the removed state, an individual is assumed to be immune to the infection. The SIR model has laid the foundation for the development of epidemic models, and has been extended to a handful of variants to reflect more specific and realistic scenarios. Some classical examples are \begin{enumerate*}[label=\emph{\roman*)}]
	\item the SIRS model that considers the immunity gained in the recovery process as temporary and allows removed individuals to become susceptible again~\cite{CS1978generalization},
	\item the SI model that neglects recovery~\cite{PV2001epidemic},
	\item the SIS model that neglects the immunity but assumes recovered individuals to become susceptible again~\cite{Newman2005threshold}, and
	\item the SEIR model that introduces an additional state E (exposed) to reflect the incubation period of the infection~\cite{Anderson1991infectious}.
\end{enumerate*}

In the context of information diffusion in social networks, it shall be taken into account that the diffusion process can relate to various factors such as time, relation strength, information content, social factors, network structure, etc. Variants have been proposed to extend classical epidemic models, aiming to reflect these correlations, such like the works in~\cite{WYX+2014seir,XLX2013research,Ding2015research,FHL+2015competing,WLJ+2015esis}. 

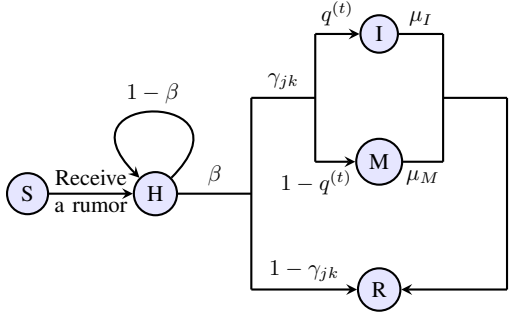
\begin{figure}[!htbp]
	\centering
	\begin{tikzpicture}[scale=0.8, transform shape, node distance=2cm,>=stealth,thick]
		\node (S) [draw, circle, fill=blue!10] {S};
		\node (H) [draw, circle, fill=blue!10, right of=S] {H};
		\node (r1) [coordinate, node distance=1.5cm, right of=H] {};
		\node (r1u) [coordinate, node distance=1.5cm, above of=r1] {};
		\node (r1d) [coordinate, node distance=1.5cm, below of=r1] {};
		\node (r2) [coordinate, right of=r1u, node distance=1cm] {};
		\node (r2u) [coordinate, above of=r2, node distance=1cm] {};
		\node (r2d) [coordinate, below of=r2, node distance=1cm] {};
		\node (I) [draw, circle, fill=blue!10, right of = r2u, node distance=1cm] {I};
		\node (M) [draw, circle, fill=blue!10, right of = r2d, node distance=1cm] {M};
		\node (r3u) [coordinate, right of=I, node distance=1cm] {};
		\node (r3d) [coordinate, right of=M, node distance=1cm] {};
		\node (r3) [coordinate, below of=r3u, node distance=1cm] {};
		\node (r4) [coordinate, right of=r3, node distance=1cm] {};
		\node (r5) [coordinate, below of = r4, node distance=3cm] {};
		\node (R) [draw, circle, fill=blue!10, below of=M] {R};
		\path[->] 	(S) edge node[midway, above] {Receive} node[midway, below] {a rumor}(H) 
					(H) edge[loop above, out=45, in=135, looseness=10] node[midway, left, xshift=5mm, yshift=3mm] {$1-\beta$} (H)
					(r1d) edge node[midway, above] {$1-\gamma_{jk}$}(R)
					(r2u) edge node[midway, above] {$q^{(t)}$} (I)
					(r2d) edge node[midway, below, xshift=-3mm] {$1-q^{(t)}$}(M)
					(r5) edge (R);
		\path[-] 	(H) edge node[midway, above] {$\beta$} (r1)
					(r1u) edge (r1d)
					(r1u) edge node[midway, above] {$\gamma_{jk}$} (r2)
					(r2u) edge (r2d)
					(I) edge node[midway, above] {$\mu_I$} (r3u)
					(M) edge node[midway, below] {$\mu_M$} (r3d)
					(r3u) edge (r3d)
					(r3) edge (r4)
					(r4) edge (r5);
	\end{tikzpicture}
	\caption{Conventional SHIMR model for an $\left<i,j\right>$-node}
	\label{fig:shimr}
\end{figure}

Another phenomenon that distinguishes the information diffusion process from the epidemic spread process is the individual's freedom in decision making. When exposed to a certain information, an individual is not destined to forward it to others. It may also simply remain silent about this information, or disprove and publicly refute it instead (known as the ``anti-rumor''). This has been considered in the SHIMR model~\cite{LWO2022rumor}, which extends the SIR model with two additional states of H (hesitant) and M (mitigated), as shown in Fig.~\ref{fig:shimr}. After receiving a rumor, in each period an individual has a probability $1-\beta$ to remain hesitant, and a probability $\beta$ to make its decision from three different options: to forward the rumor, to refute the rumor, or to remain silent. In each of the two former cases, the individual will eventually lose its interest in this rumor and become silent after a random period of time. The probability $\gamma_{jk}$ of remaining silent depends on the connection degrees $\left<j,k\right>$ of the individual to the other individuals in the network, and the tendency of selecting between approval and disproval is determined by the probability $q^{(t)}$, which is a function of the time $t$.

\section{The Extended SHIMR Model}\label{sec:model}
The conventional SHIMR model is able to capture the individual's decision making process after receiving a rumor, but is overgeneric or oversimplified in its specification to the transition probabilities. Inspired by the knowledge in social psychology, we propose to extend the SHIMR model, aiming to encompass the dynamic decision making process of each individual upon the social environment around.
\subsection{Correlated Rumors, Value, Opinion, and States}\label{subsec:correlated_rumors}
First, instead of focusing on the speading of one single ruomr, we consider the case where multiple intercorrelated rumors, which share a same subject or ideology, are successively released to the network. For a network with node set $\mathcal{N}$, we consider each node $n\in\mathcal{N}$ to be featured with its \emph{opinion} $o_n\in[-1,1]$ on this subject (e.g., $o_n=-1$ for radically libral, $o_n=1$ for radically conservative, and $o_n=0$ for neutral). Similarly, each rumor $k$ spreading in the network is associated with its \emph{value} $v_k\in[-1,1]$.

We consider all agents (nodes) in the network to be categorized into two sets: the influencers $\mathcal{N}_\mathrm{I}$ and normal agents $\mathcal{N}_\mathrm{N}$. Influeners generate rumors and spread them to the network, but never receive rumors from others. They have consistent opinions that do not diffuse, and all rumors generated by an influencer have the same fixed value that is consistent with the influencer's opinion. Normal agents, on contrary, receive rumors, make decisions based on their own beliefs and the social environment, and forward/refute the rumors.

\begin{figure}[!htpb]
	\centering
	\begin{tikzpicture}[scale=0.8, transform shape, node distance=2cm,>=stealth,thick]
		\node (S) [draw, circle, fill=blue!10] {S};
		\node (H) [draw, circle, fill=blue!10, right of=S] {H};
		\node (r1) [coordinate, node distance=1.5cm, right of=H] {};
		\node (R) [draw, circle, fill=blue!10, right of = r1] {R};
		\node (I) [draw, circle, fill=blue!10, above of =R] {I};
		\node (M) [draw, circle, fill=blue!10, below of=R] {M};
		\node (r1u) [coordinate, node distance=1cm, above of=r1] {};
		\node (r1d) [coordinate, node distance=1cm, below of=r1] {};
		\node (r2) [coordinate, right of=r1u, node distance=1cm] {};
		\node (r2u) [coordinate, above of=r2, node distance=0.5cm] {};
		\node (r2d) [coordinate, below of=r2, node distance=0.5cm] {};
		\node (r3) [coordinate, right of=r1d, node distance=1cm] {};
		\node (r3u) [coordinate, above of=r3, node distance=0.5cm] {};
		\node (r3d) [coordinate, below of=r3, node distance=0.5cm] {};
		\node (r4) [coordinate, right of=R, node distance=1cm] {};
		\node (r4u) [coordinate, right of=I, node distance=1cm] {};
		\node (r4d) [coordinate, right of=M, node distance=1cm] {};
		\path[->] 	(S) edge node[midway, above] {$\alpha_{n,k}$} (H)
					(S) edge[loop left, out=135, in=215, looseness=5] node[midway, above, xshift=3mm, yshift=3mm] {$1-\alpha_{n,k}$} (S)
					(H) edge[loop above, out=45, in=135, looseness=5] node[midway, left, xshift=5mm, yshift=3mm] {$1-\beta_n$} (H)
					(r2u) edge node[midway, below, xshift = 4mm] {$\gamma_{n,k}$} (I)
					(r2d) edge node[midway, above, xshift = 4mm] {$1-\gamma_{n,k}$} (R)
					(r3u) edge node[midway, below, xshift = 4mm] {$1-\gamma_{n,k}$} (R)
					(r3d) edge node[midway, above, xshift = 4mm] {$\gamma_{n,k}$}(M)
					(r4) edge (R)
					(I) edge[loop above, out=45, in=135, looseness=5] node[midway, above] {$1-\mu_{n,k}$} (I)
					(M) edge[loop below, out=315, in=215, looseness=5] node[midway, below] {$1-\mu_{n,k}$} (M)
					;
		\path[-] 	(H) edge node[midway, above] {$\beta_n$} (r1)
					(r1u) edge (r1d)
					(r1u) edge node[midway, above, xshift = -3mm] {$q_{n,k}$} (r2)
					(r1d) edge node[midway, below, xshift = -3mm] {$1-q_{n,k}$} (r3)
					(r2u) edge (r2d)
					(r3u) edge (r3d)
					(I) edge node[midway, above, xshift = 4mm] {$\mu_{n,k}$} (r4u)
					(M) edge node[midway, below, xshift = 4mm] {$\mu_{n,k}$}  (r4d)
					(r4u) edge (r4d)
					;
	\end{tikzpicture}
	\caption{Proposed model for a normal agent $n$ upon rumor $k$}
	\label{fig:shimr_ext}
\end{figure}
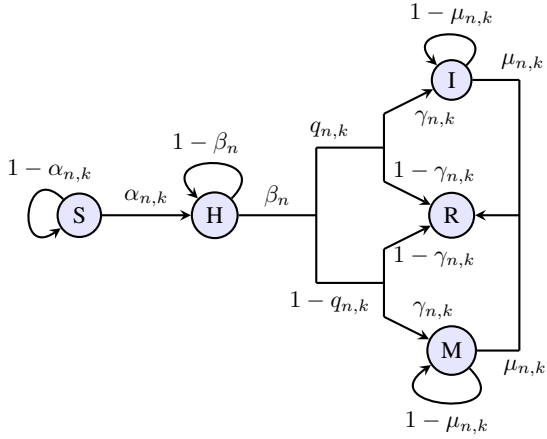

Compared to conventional epidemic models, our model considers each node to have an individual status for every rumor. The decision-making process of a normal agent $n\in\mathcal{N}_\mathrm{N}$ upon a rumor $k$ is illustrated in Fig.~\ref{fig:shimr_ext}: $n$ is initialized in the state $s_{n,k}^{(0)}=\text{S}$ for every $k$, and in every round $t$ it has a probability $\alpha_{n,k}^{(t)}$ to receive the rumor $k$ from the network and transit into the state H. In the state H, every round $n$ has a probability $\beta_n^{(t)}$ to make its decision regarding the rumor. The decision is by a chance of $q_{n,k}^{(t)}$ to approve the rumor $k$, and by $1-q_{n,k}^{(t)}$ to disprove. Subsequently, the agent will make a second decision on if it will publicly share its opinion in the network. It has a probability of $\gamma_{n,k}^{(t)}$ to forward/refute the rumor and transit into the state I/M, in cases of approval/disproval, respectively, and a probability of $1-\gamma_{n,k}^{(t)}$ to remain silent and directly transit into the state R. A normal agent in state I or M will have a probability of $\mu_{n,k}^{(t)}$ to lose interest in the rumor $k$ and transit into the state R in every round.

\subsection{Non-Linear Diffusion of Opinion}\label{subsec:nonlinear_difussion}
Upon deciding to approve or disprove a rumor $k$, an agent $n$ will have its opinion $o_n$ diffusing, either towards the rumor's value $v_k$ in the case of approval, or apart from it otherwise. It shall be noticed that the opinion is bounded in $[-1,1]$, and the diffusion process shall be non-linear: it is easier to change a neutral or mild opinion than to shake a radical one. To this end, we define a tangential opinion index $\phi_n^{(t)} \triangleq \tan\left(\frac{\pi}{2} o_n^{(t)}\right)$ to implement the following non-linear diffusion:
\begin{align}
	\Delta \phi_{n,k}^{(t)}&= \Lambda d_{n,k}\mathrm{sgn}\left(v_k-o_n^{(t)}\right),\label{eq:opinion_index_change}\\
	o_n(t+1) &= \frac{2}{\pi}\arctan\left(\rho\phi_n^{(t)}+\sum\limits_{k\in\mathcal{K}^{(t)}}\Delta\phi_{n,k}^{(t)}\right),\label{eq:opinion_update}
\end{align}
where $\Lambda\in\mathbb{R}^+$ is the \emph{influence factor}, and $d_{n,k}^{(t)}$ is the indicator for agent $n$ making its decision about rumor $k$ in round $t$:
\begin{equation}
	d_{n,k}^{(t)}=\begin{cases}
		1, & s_{n,k}^{(t-1)}=\text{H} \land s_{n,k}^{(t)}=\text{I}\\
		-1, & s_{n,k}^{(t-1)}=\text{H} \land s_{n,k}^{(t)}=\text{M}\\
		0, & \text{otherwise}
	\end{cases},
\end{equation}
$\mathcal{K}$ is the set of all rumors in network, and $\rho\in(0,1)$ the \emph{memory factor} that scales agents' opinion consistency.

\subsection{Social Weight and Homophily}\label{subsec:social_weight}
While the conventional SHIMR model considers all connections between different nodes to be equally important, we consider each directional social connection to be associated with an individual \emph{social weight}. We define the weight matrix $\mathbf{W}_{N\times N}$ where each element $w_{m,n}\in[0,1]$ is the social weight of node $m$ for node $n$, a larger $w_{m,n}$ indicating a stronger influence of $m$ on $n$. More specifically, when $m$ is publishing/forwarding a rumor in the network, $w_{m,n}^{(t)}$ is the probability that $n$ receives it from $m$ in round $t$.

Social psychology has repeatedly confirmed the well-known phenomenon of \emph{homophily}~\cite{EBK+2022homophily}, i.e. the tendency of individuals to associate with similar ones rather than dissimilar ones. To reflect this, we let $\mathbf{W}$ be updated at the beginning of each round $t$, regarding the opinions of the nodes in the network:
\begin{equation}
		w_{m,n}^{(t)}=\label{eq:social_weight_update}
		\begin{cases}
			1-e^{\eta\left(\left\vert o_m-o_n\right\vert-O\right)}\left[1-w_{m,n}^{(t-1)}\right],& \left\vert o_m-o_n\right\vert\leqslant O\\
			e^{\eta\left(O-\left\vert o_m-o_n\right\vert\right)}w_{m,n}^{(t-1)},& \text{otherwise}
		\end{cases},
\end{equation}
where $\eta$ is the \emph{crowd exponent} that determines the strength of homophily, and $O$ is the \emph{concensus threshold} of opinion difference to consider two individuals similar with each other. 
\subsection{Exposure to Rumors}
How likely a node receives a rumor depends on two factors:
\begin{enumerate*}[label=\emph{\roman*)}]
	\item the number of its social relatives discussing this rumor (or its anti-rumor), and
	\item the social weight of each of these relatives.
\end{enumerate*}
Given a certain node $n$ in a certain round $t$, we consider its receptions of the same rumor $k$ from different other nodes $m\neq n$ to be events independent from each other, thus, the probability of $n$ being exposed to rumor $k$ is outlined by
\begin{equation}\label{eq:exposure}
	\begin{split}
		&\alpha_{n,k}^{(t)}\triangleq\text{Prob}\left[s_{n,k}^{(t)}=\text{H}\vert s_{n,k}{(t-1)}=\text{S}\right]\\
		=&1-\prod\limits_{m\in\mathcal{N}^+_{n,k}{(t-1)}}\left[1-w_{m,n}^{(t)}\right]
	\end{split},
\end{equation}
where $\mathcal{N}^+_{n,k}{(t)}=\left\{m\in\mathcal{N}: m\neq n, s_{m,k}^{(t)}\in\{\text{I,M}\}\right\}$ is the set of nodes (other than $n$) discussing rumor $k$ in round $t$.

\subsection{Sociocognitive Decision Making}\label{subsec:decision_making}
Once exposed to a rumor, an agent will take time to assess the creditability of the rumor and eventually make its decision. The speed of this process can be determined by two factors: the agent's own opinion and the social consensus it perceives. On the one hand, people with more extreme ideological positions often display higher tendency to form quicker judgments are more resistant to processing new information thoroughly, which is known as the phenomenon of \emph{cognitive rigidity}~\cite{Zmigrod2020role} or \emph{need for cognitive closure}~\cite{WK1994individual}. On the other hand, it triggers deeper (and slower) information processing in decision making when people encounters more diverse viewpoints, which is supported by the \emph{conflict elaboration theory}~\cite{RMR+2014information}. To capture this, we define the decision-making probability of a normal agent $n$ in round $t$ as
\begin{equation}\label{eq:decision_making}
	\beta_n=\max\left\{\vert o_n\vert(1-\sigma_{n,k}),\beta_\text{min}\right\}.
\end{equation}
Here, $\beta_\text{min}$ is the minimum decision chance, $\sigma_{n,k}\in[0,1]$ the deviation of neighbor decisions on rumor $k$ perceived by $n$, weighted regarding the social weights of its neighbor nodes:
\begin{align}
	\sigma_{n,k}&=\sqrt{\left.\sum\limits_{m\in\mathcal{N}^+_{n,k}}w_{m,n}(i_{m,k}-I_{n,k})^2\right/{\sum\limits_{m\in\mathcal{N}^+_{n,k}}w_{m,n}}},\\
	i_{n,k}&=\begin{cases}
		1, &s_{n,k}=\text{I}\\
		-1, &s_{n,k}=\text{M}\\
		0, &\text{otherwise}
	\end{cases},\\
	I_{n,k}&=\frac{1}{N-1}\sum\limits_{m\in\mathcal{N}^+_{n,k}}i_{m,k}.
\end{align}

Furthermore, the agent's specific decision between approval and disproval is jointly determined by the its own opinion $o_n$ and the rumor's value $v_k$. It has been revealed by multiple evidences that when people are exposed to information that is against their own beliefs, they tend to enhance their own beliefs rather than to change them, known as the \emph{backfire effect}~\cite{BAB+2018exposure,WP2019elusive}. We model this phenomenon by specifying the probability of approval as a linear function of the difference between the agent's opinion and the rumor's value:
\begin{equation}\label{eq:side_taking}
	q_{n,k}=1-\frac{\left\vert v_k-o_n\right\vert}{2}.
\end{equation}

\subsection{Self-Presentation-Driven Expression}\label{subsec:expression}
Another well-tested phenomenon in social psychology is selective \emph{self-presentation}, which states that people tend to present themselves consistently with their beliefs and values. The same effect is also observed in online social media~\cite{Hollenbaugh2021self}. Therefore, we consider the willingness of an agent $n$ to participate in discussion, i.e., its probability to spread or publicly refute a rumor $k$, as a function of its opinion $o_n$, the rumor's value $v_k$, and its decision:
\begin{equation}\label{eq:self_presentation}
	\gamma_{n,k}=\exp\left(-\Gamma \left\vert o_n-i_{n,k}v_k\right\vert\right),
\end{equation}
where $\Gamma>0$ is the \emph{silence exponent} that measures the extent of the self-presentation effect. 

\subsection{Popularity-Based Loss of Interest}\label{subsec:loss_interest}
Last but not least, the process of an agent losing interest in a rumor is jointly determined by the agent-perceived popularity of the rumor and the agent's own willingness of discussion. Taking both factors into account, we model the probability of an agent $n$ losing interest in rumor $k$ in round $t$ as
\begin{equation}\label{eq:loss_interest}
	\mu_{n,k}= 1-\xi\alpha_{n,k}\gamma_{n,k},
\end{equation}
where the \emph{trend factor} $\xi\in(0,1]$ scales the interest consistency.

\subsection{Overall Process}\label{subsec:overall_process}
Taking into account the mechanisms described above in Sec.~\ref{subsec:correlated_rumors}--\ref{subsec:loss_interest}, the overall process of opinion diffusion in the network is summarized in Algorithm~\ref{alg:overall_process}.
\begin{algorithm}[!htpb]
	\caption{Overall opinion diffusion process}
	\label{alg:overall_process}
	\scriptsize
	\DontPrintSemicolon
	Specify: $\mathcal{N}_\mathrm{I}, \mathcal{N}_\mathrm{N}, T, R, \Lambda, \eta, O, \rho, \beta_\text{min}, \xi$\; 
	Initialize: $\mathcal{K}=\emptyset, \mathbf{W}, \mathbf{o}$\;
	\For(\tcp*[f]{Limited rounds of iteration}){$t\in\{1,2,\dots T\}$}{
		Update $\mathbf{o}$ according to Eq.~\eqref{eq:opinion_update}\;
		Update $\mathbf{W}$ according to Eq.~\eqref{eq:social_weight_update}\;
		\ForEach{$n\in\mathcal{N}_\mathrm{I}$}{
			Generate a new rumor $k$ with value $v_k=o_n$\;
			$\mathcal{K}\gets \mathcal{K}\cup\{k\}$\tcp*[f]{Rumor generation}
		}
		\ForEach{$k\in\mathcal{K}$}{
			\ForEach{$n\in\mathcal{N}_\mathrm{N}$}{
				Update model parameters according to Eqs.~\eqref{eq:exposure}--\eqref{eq:loss_interest}\;
				Update $s_{n,k}$ according to Fig.~\ref{fig:shimr_ext}\;
				Calculate $\Delta\phi_{n,k}$ according to Eq.~\eqref{eq:opinion_index_change}\;
			}
			\If(){$s_{n,k}=\text{R},\quad\forall n\in\mathcal{N}_\mathrm{N}$}{$\mathcal{K}\gets \mathcal{K}\backslash\{k\}$\tcp*[f]{Remove expired rumors}}
		}
	}
	\Return{$\mathbf{W},\mathbf{o}$}
\end{algorithm}
\section{Numerical Simulations}\label{sec:simualtion}
\subsection{Simulation Setup}\label{subsec:setup}
To verify if our proposed model can effectively capture the sophisticated behavior of social networks such like the echo chamber effect, and to explore its sensitivity to different model parameters, we carried out a numerical simulation campaign. The basic simulation setup is summarized in Tab.~\ref{tab:setup}. 
\begin{table}[!htbp]
	\centering
	\caption{Simulation Setup}
	\label{tab:setup}
	\begin{tabular}{>{\cellcolor{white}}m{0.2cm} | m{1.3cm} l m{4.5cm}}
		\toprule[2px]
		&\textbf{Parameter}&\textbf{Value}&\textbf{Remark}\\
		\midrule[1px]
		
		\rowcolor{gray!20}
		&	$N$	&	100	&	Number of agents\\
		&	$T$	&	150 &	Rounds of iteration per run\\
		
		\rowcolor{gray!20}
		\multirow{-3}{*}{\rotatebox{90}{\textbf{System}}}	&	$R$	&	500	&	Number of runs per Monte-Carlo test\\
		\midrule[1px]
		&	$\Lambda$	&	1	&	Influence factor\\
		
		\rowcolor{gray!20}
		&	$\rho$	&	0.5	&	Memory factor, see Eq.~\eqref{eq:opinion_update}\\
		&	$\beta_\text{min}$	&	0.01	&	Minimal decision chance, see Eq.~\eqref{eq:decision_making}\\

		\rowcolor{gray!20}
		\multirow{-4}{*}{\rotatebox{90}{\textbf{Model}}}	&	$\xi$	&	0.8	&	Trend factor, see Eq.~\eqref{eq:loss_interest}\\
		
		\bottomrule[2px]
	\end{tabular}
\end{table}

\begin{figure}[!htbp]
	\centering
	\begin{subfigure}[t]{.45\linewidth}
		\centering
		\includegraphics[height=.77\linewidth, trim=46 48 125 20, clip]{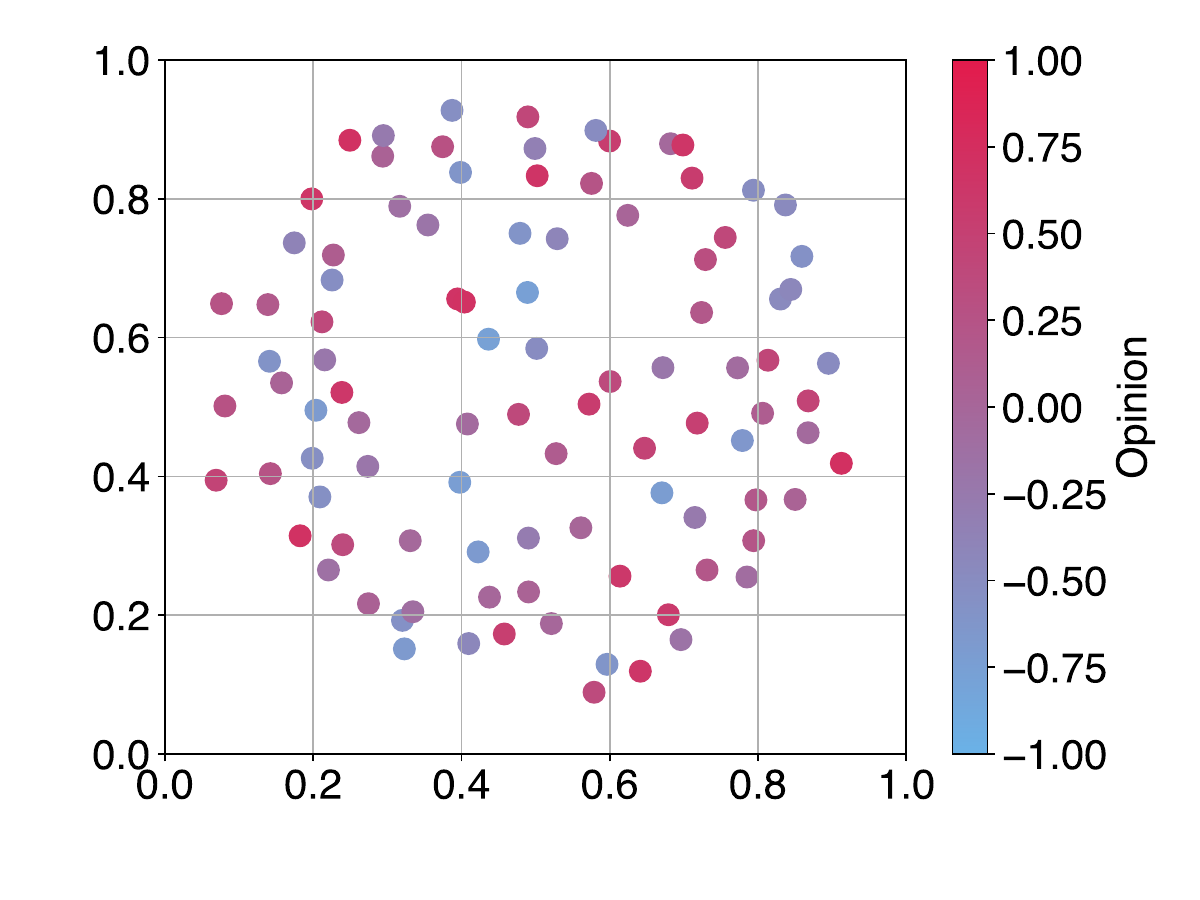}
		\caption{}
		\label{subfig:echo_chamber_init}
	\end{subfigure}
	\begin{subfigure}[t]{.53\linewidth}
		\centering
		\includegraphics[height=.654\linewidth, trim=46 48 10 20, clip]{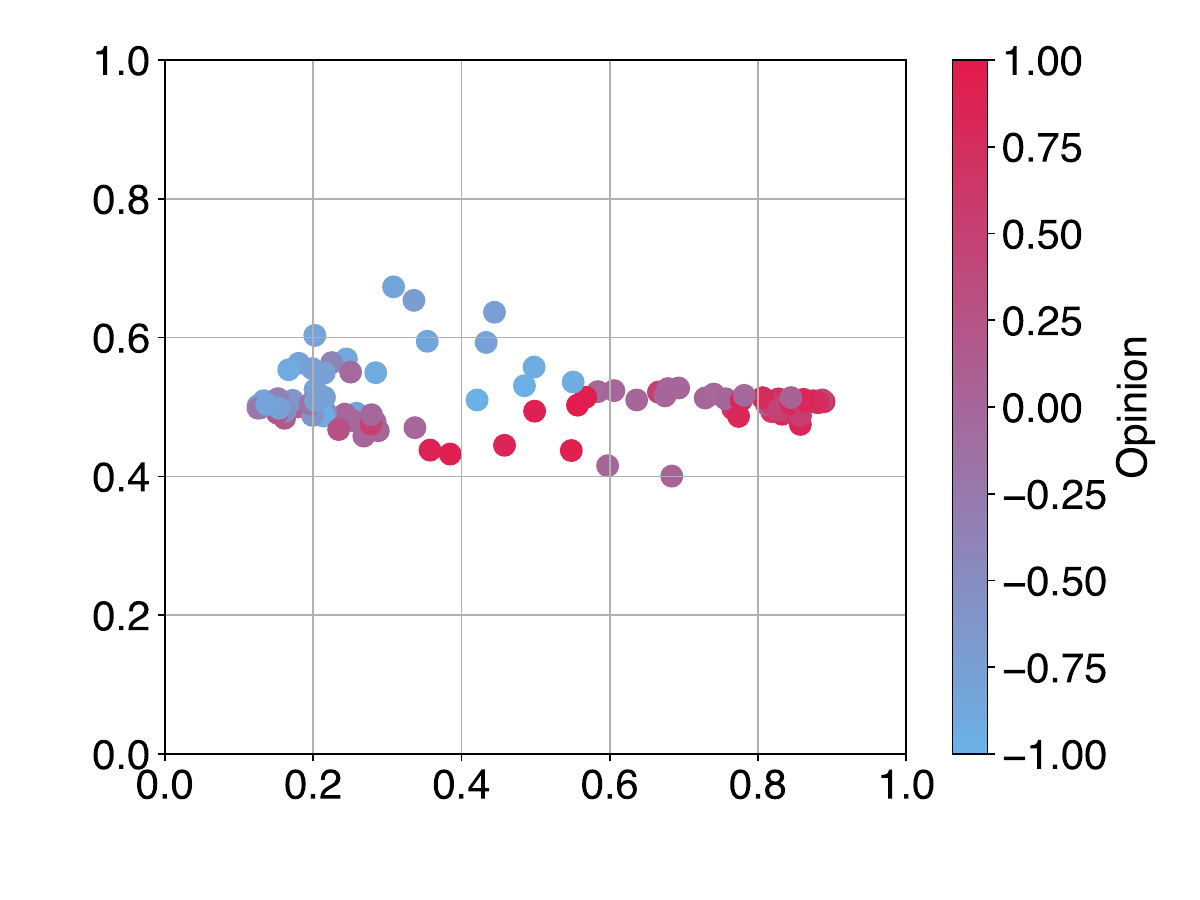}
		\caption{}
		\label{subfig:echo_chamber_example}
	\end{subfigure}
	\caption{Opinions of normal agents and the social relations among them, \subref{subfig:echo_chamber_init} before and\subref{subfig:echo_chamber_example} after 150 rounds of diffusion. The visualized distance between each two nodes is inversely proportional to the social weight between them.}
	\label{fig:echo_chamber}
\end{figure}

\subsection{Observing the Echo Chamber Effect}
First, a case study was conducted to observe the echo chamber effect. We set the system parameters $(\eta,O,\Gamma)=(0.1,1,1)$.
All agents were first initialized with Gaussian random opinion indices $\sim\mathcal{N}(0,1)$, and the weight factors between each pair of agents is initialized to uniformly distributed random values $\sim\mathcal{U}(0,1)$. 
Then, two random agents were selected to be the influencers opposite to each other, granted with static opinions $-1$ and $1$, respectively. Each influencer was set to periodically generate rumors at a rate of one rumor per round. After 150 iterations, observed are a significant polarization of opinions and a clear division of the community, highly correlated to each other, exhibiting the echo chamber effect, as shown in Fig.~\ref{fig:echo_chamber}.

\begin{figure*}
	\centering
	\begin{subfigure}[t]{.749\linewidth}
		\centering
		\fbox{\includegraphics[width=\linewidth, trim=12 12 12 10, clip]{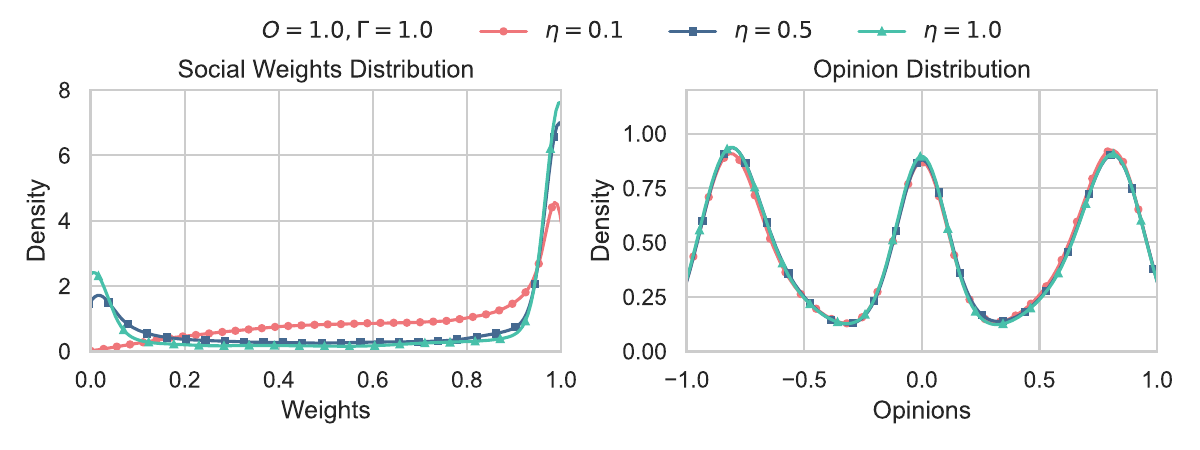}}
		\caption{Comparison w.r.t. different values of the crowd factor $\eta$}
		\label{subfig:sensitivity_test_eta}
	\end{subfigure}
	\begin{subfigure}[t]{.75\linewidth}
		\centering
		\fbox{\includegraphics[width=\linewidth, trim=12 12 12 8, clip]{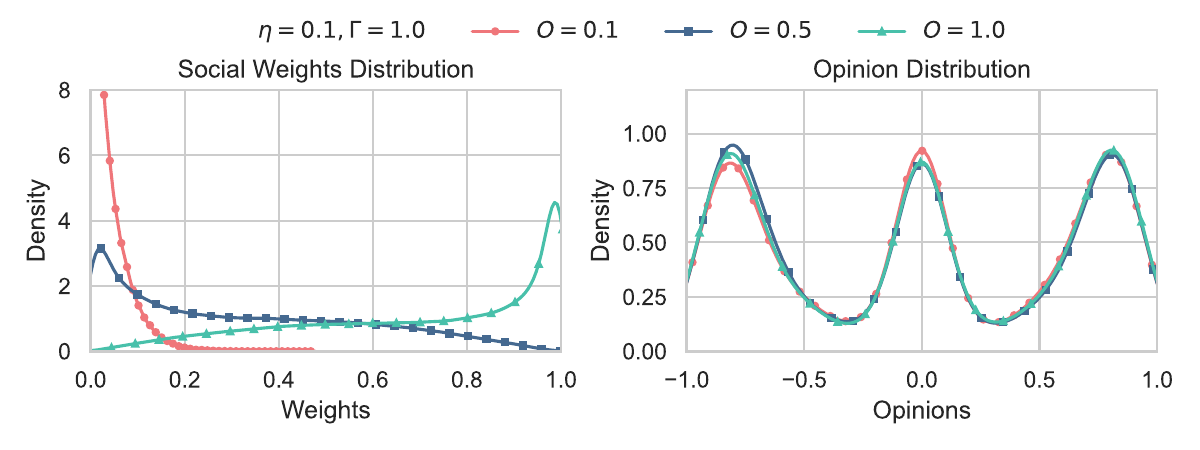}}
		\caption{Comparison w.r.t. different values of the concensus threshold $O$}
		\label{subfig:sensitivity_test_O}
	\end{subfigure}
	\begin{subfigure}[t]{.75\linewidth}
		\centering
		\fbox{\includegraphics[width=\linewidth, trim=12 12 12 8, clip]{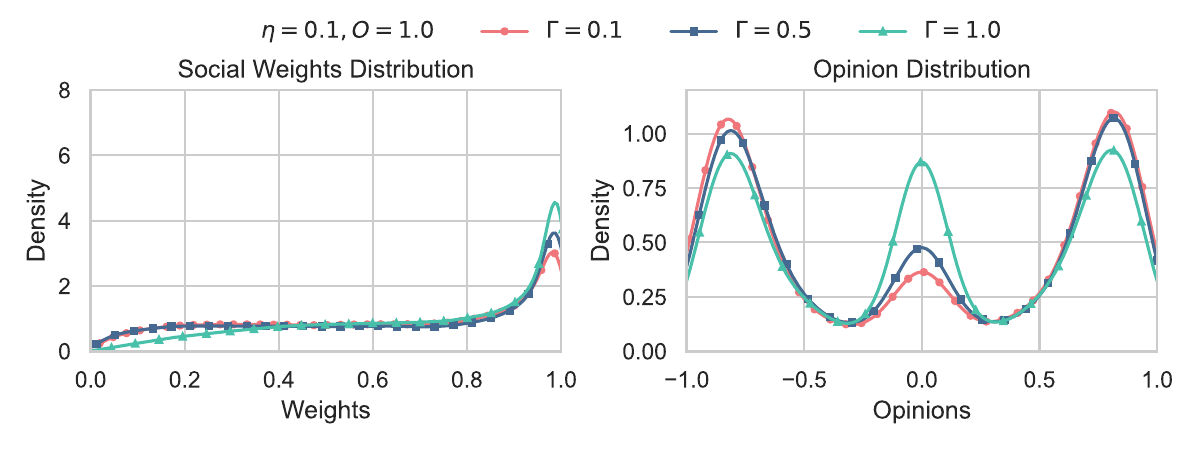}}
		\caption{Comparison w.r.t. different values of the silence exponent $\Gamma$}
		\label{subfig:sensitivity_test_Gamma}
	\end{subfigure}
	\begin{subfigure}[t]{.75\linewidth}
		\centering
		\fbox{\includegraphics[width=\linewidth, trim=12 12 12 10, clip]{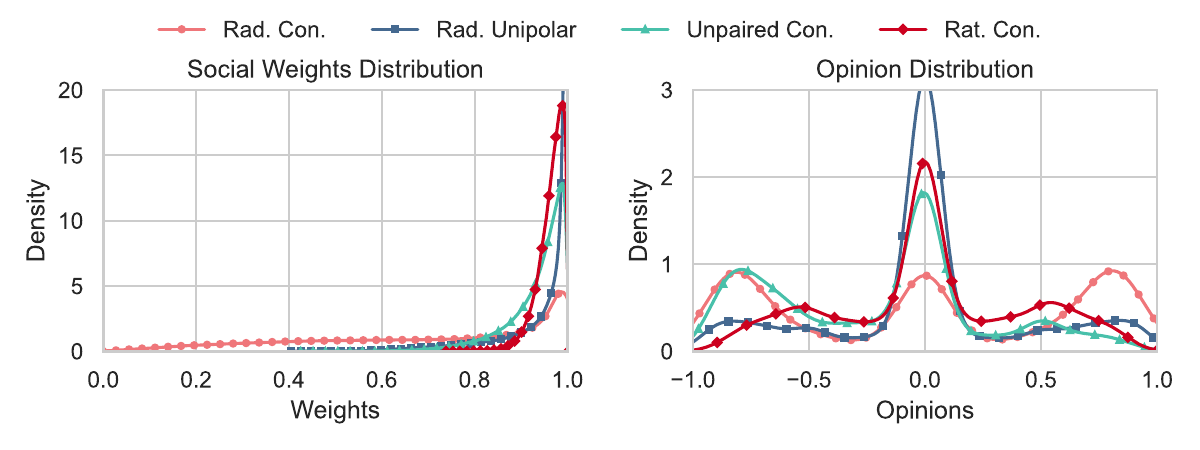}}
		\caption{Comparison w.r.t. different influencer settings}
		\label{subfig:influencer_impact_test}
	\end{subfigure}
	\caption{Distributions of social weight and opinion after 150 rounds of diffusion.}
	\label{fig:sensitivity_tests}
\end{figure*}

A quantitative assessment was conducted by Monte Carlo tests under the same setup. We repeated the same simulation 100 times for the statistical distributions of the social connection weights and agent opinions after 150 diffusion rounds, which are illustrated with coral red curves in Fig.~\ref{subfig:sensitivity_test_eta}. In the social weights distribution, we observe a single peak at the higher end and a long tail at the lower, implying that \emph{the agents are organizing themselves into tightly connected groups}. In the opinion distribution, we see three symmetric clusters, one biased towards each of the two influencers, and one in the middle, indicating that \emph{the agents are polarized} into two groups with extreme opinions, and a remaining group of neutral agents. More specifically, by co-analyzing the social weight between each pair of agents and their opinion differences, we came to a correlation coefficient of -0.555, exhibiting a strong correlation between them. In conclusion, the echo chamber effect is significantly captured by the simulation results.

\subsection{Sensitivity Tests}
Subsequently, to explore the impact of model parameters, we conducted a series of sensitivity tests by repeating the Monte Carlo simulations with different values of the model parameters. The distributions of resulted social weights and opinions are illustrated in Figs.~\ref{subfig:sensitivity_test_eta}--\subref{subfig:sensitivity_test_Gamma}, and the correlation coefficients between the social weights and opinion differences are summarized in Tab.~\ref{tab:sensitivity_test_results}, where the baseline setup is highlighted in gray cells. We can observe that the echo chamber effect can be significantly enhanced by increasing the crowd exponent $\eta$ or reducing the silence exponent $\Gamma$. The role of the concensus threshold $O$ is more subtle: it has little impact on the opinion, but significantly affects the social relation: a lower $O$ implies a stricter standard for agreement and will lead to a signifcant distancing of the agents from each other.

\begin{table}[!htbp]
	\centering
	\caption{Weight-opinion correlation upon model parameters}
	\label{tab:sensitivity_test_results}
	\begin{tabular}{c | c c c}
		\toprule[2px]
		\multirow{2}{*}{$(O,\Gamma)=(1.0,1.0)$}	&	$\cellcolor{gray!20}\eta=0.1$	&	$\eta=0.5$	& $\eta=1.0$\\
		&	\cellcolor{gray!20}-0.555	&	-0.774	& -0.819\\
		\midrule[1px]
		\multirow{2}{*}{$(\eta,\Gamma)=(0.1,1.0)$}	&	$O=0.1$	&	$O=0.5$	& \cellcolor{gray!20}$O=1.0$\\
		&	-0.428	&	-0.535	& \cellcolor{gray!20}-0.555\\
		\midrule[1px]
		\multirow{2}{*}{$(\eta,O)=(0.1,1.0)$}	&	$\Gamma=0.1$	&	$\Gamma=0.5$	& \cellcolor{gray!20}$\Gamma=1.0$\\
								&	-0.608	&	-0.612	& \cellcolor{gray!20}-0.555\\
		\bottomrule[2px]
	\end{tabular}
\end{table}

At last, to investigate the impact of influencer nodes, we repeated the test under different influencer settings. In addition to the baseline with two radical influencers opposing each other with opinions $(-1,1)$, which we call \emph{Radical Controversy}, we considered three other setups: \begin{enumerate*}[label=\emph{\roman*)}]
	\item \emph{Radical Unipolar}, with only one radical influencer of opinion $-1$;
	\item \emph{Unpaired Controversy}, with two influencers of opposite opinions $(-1,0.3)$; and
	\item \emph{Rational Controversy}, with two influencers of opposite but mild opinions $(-0.3,0.3)$.
\end{enumerate*}
The model parameters were fixed to $(\eta,O,\Gamma)=(0.1,1,1)$. The results are shown Fig.~\ref{subfig:influencer_impact_test} and Tab.~\ref{tab:influencer_impact}. Meeting our expectation, the \emph{Radical Controversy} setup leads to the most significant opinion polarization and the most split social network. In contrast, the \emph{Rational Controversy} setup results in the most moderate polarization and a steady social network. The most interesting results, however, are observed with the other two setups: 
The \emph{Radical Unipolar}, where the influencer power is asymmetric, does not end up with a skewed opinion distribution, but a symmetric and even least polarized one. It also leads to the most tightly connected social network. Being counter-intuitive, it can be explained by the fact that a unipolar radical influencer is initially far distanced from the majority of the population, and thus more declining than attracting people, unless this effect is balanced by a similarly radical opponent. This result acknowledges the widely-suggested strategy of ``Don't feed the trolls.''
The \emph{Unpaired Controversy} setup, though resulting in a skewed opinion distribution with a significantly minus-biased cluster, yet leads to less opinion polarization and a more steadily connected social network when compared to the \emph{Radical Controversy}, acknowledging another clich\'e ``Logic disarms lunacy.''

\begin{table}[!htbp]
	\centering
	\caption{Weight-opinion correlation upon influencers}
	\label{tab:influencer_impact}
	\begin{tabular}{c c c c}
		\toprule[2px]
		\cellcolor{gray!20}\textbf{Rad. Con.}	&	\textbf{Rad. Unipolar}	& \textbf{Unpaired Con.}	& \textbf{Rat. Con.}\\
		\midrule[1px]
		\cellcolor{gray!20}-0.555	&	-0.307	&	-0.275	& -0.252\\
		\bottomrule[2px]
	\end{tabular}
\end{table}

\section{Conclusion}\label{sec:conclusion}
In this paper, we proposed an extended SHIMR model to capture the dynamics of successive opinion diffusion in social networks. By incorporating mechanisms such as non-linear opinion diffusion, homophily-driven social weights, and sociocognitive decision-making, the model reflects complex phenomena like the echo chamber effect. Numerical simulations validated its ability to capture opinion polarization and social network evolution. Sensitivity tests revealed the impact of key parameters, while experiments with different influencer setups provided insights into strategies for mitigating polarization. Future work may explore real-world data validation and further extensions to account for additional social factors.

\section*{Acknowledgment}
This work is supported by the German Federal Ministry of Education \& Research in the project Open6GHub (16KISK003K / 16KISK004). B. Han (bin.han@rptu.de) is the corresponding author.

\bibliographystyle{IEEEtran}
\bibliography{references}

\end{document}

%% file: glossary.tex
\makeglossaries
\newacronym{6g}{6G}{Sixth Generation}
\newacronym{awgn}{AWGN}{additive white Gaussian noise}
\newacronym{bcd}{BCD}{block coordinate descent}
\newacronym{bri}{BRI}{biregular irreducible}
\newacronym{cp}{CP}{control plane}
\newacronym{csi}{CSI}{channel state information}
\newacronym{csit}{CSIT}{channel state information at transmitter}
\newacronym{dft-s-ofdm}{DFT-s-OFDM}{Discrete Fourier Transform-spread-OFDM}
\newacronym{fbl}{FBL}{finite blocklength}
\newacronym{gan}{GAN}{generative adversarial network}
\newacronym{ibl}{IBL}{infinite blocklength}
\newacronym{lfp}{LFP}{leakage-failure probability}
\newacronym{mcs}{MCS}{modulation and coding scheme}
\newacronym{mimo}{MIMO}{multi-input multi-output}
\newacronym{mm}{MM}{Minorize-Maximization}
\newacronym{noma}{NOMA}{non-orthogonal multi-access}
\newacronym{nom}{NOM}{non-orthogonal multiplexing}
\newacronym{ofdm}{OFDM}{orthogonal frequency-division multiplexing}
\newacronym{ofdma}{OFDMA}{orthogonal frequency-division multiple access}
\newacronym{oma}{OMA}{orthogonal multiple access}
\newacronym{papr}{PAPR}{Peak-to-Average Power Ratio}
\newacronym{pdf}{PDF}{probability density function}
\newacronym{per}{PER}{packet error rate}
\newacronym{phy}{PHY}{physical}
\newacronym{pld}{PLD}{physical layer deception}
\newacronym{pls}{PLS}{physical layer security}
\newacronym{prb}{PRB}{physical resource block}
\newacronym{sic}{SIC}{successive interference cancellation}
\newacronym{sinr}{SINR}{signal-to-interference-and-noise ratio}
\newacronym{snr}{SNR}{signal-to-noise ratio}
\newacronym{tdma}{TDMA}{time-division multiple access}
\newacronym{up}{UP}{user plane}
\newacronym{urllc}{URLLC}{ultra-reliable low-latency communication}

%% file: main.bbl
\begin{thebibliography}{10}
\providecommand{\url}[1]{#1}
\csname url@samestyle\endcsname
\providecommand{\newblock}{\relax}
\providecommand{\bibinfo}[2]{#2}
\providecommand{\BIBentrySTDinterwordspacing}{\spaceskip=0pt\relax}
\providecommand{\BIBentryALTinterwordstretchfactor}{4}
\providecommand{\BIBentryALTinterwordspacing}{\spaceskip=\fontdimen2\font plus
\BIBentryALTinterwordstretchfactor\fontdimen3\font minus \fontdimen4\font\relax}
\providecommand{\BIBforeignlanguage}[2]{{%
\expandafter\ifx\csname l@#1\endcsname\relax
\typeout{** WARNING: IEEEtran.bst: No hyphenation pattern has been}%
\typeout{** loaded for the language `#1'. Using the pattern for}%
\typeout{** the default language instead.}%
\else
\language=\csname l@#1\endcsname
\fi
#2}}
\providecommand{\BIBdecl}{\relax}
\BIBdecl

\bibitem{LWG+2017survey}
M.~Li \emph{et~al.}, ``A survey on information diffusion in online social networks: Models and methods,'' \emph{Information}, vol.~8, p. 118, 2017.

\bibitem{CDG+2021echo}
M.~Cinelli \emph{et~al.}, ``The echo chamber effect on social media,'' \emph{Proc. Natl. Acad. Sci. U. S. A.}, vol. 118, no.~9, p. e2023301118, 2021.

\bibitem{LWO2022rumor}
W.~Liu, J.~Wang, and Y.~Ouyang, ``Rumor transmission in online social networks under {Nash} equilibrium of a psychological decision game,'' \emph{Netw. Spat. Econ.}, vol.~22, pp. 831--854, 2022.

\bibitem{KM1927contribution}
W.~O. Kermack and A.~G. McKendrick, ``A contribution to the mathematical theory of epidemics,'' \emph{Proc. R. soc. Lond. Ser. A - Contain. Pap. Math. Phys. Character}, vol. 115, no. 772, pp. 700--721, 1927.

\bibitem{CS1978generalization}
V.~Capasso and G.~Serio, ``A generalization of the {Kermack-McKendrick} deterministic epidemic model,'' \emph{Math. Biosci.}, vol.~42, no. 1-2, pp. 43--61, 1978.

\bibitem{PV2001epidemic}
R.~Pastor-Satorras and A.~Vespignani, ``Epidemic spreading in scale-free networks,'' \emph{Phys. Rev. Lett.}, vol.~86, no.~14, p. 3200, 2001.

\bibitem{Newman2005threshold}
M.~E. Newman, ``Threshold effects for two pathogens spreading on a network,'' \emph{Phys. Rev. Lett.}, vol.~95, no.~10, p. 108701, 2005.

\bibitem{Anderson1991infectious}
R.~Anderson, \emph{Infectious Diseases of Humans: Dynamics and Control}.\hskip 1em plus 0.5em minus 0.4em\relax Oxford University Press, 1991.

\bibitem{WYX+2014seir}
C.~Wang \emph{et~al.}, ``{SEIR}-based model for the information spreading over {SNS},'' \emph{Acta Electron. Sin.}, vol.~11, pp. 2325--2330, 2014.

\bibitem{XLX2013research}
R.~Xu, H.~Li, and C.~Xing, ``Research on information dissemination model for social networking services,'' \emph{Int. J. Comput. Sci. Appl}, vol.~2, pp. 1--6, 2013.

\bibitem{Ding2015research}
X.~Ding, ``Research on propagation model of public opinion topics based on {SCIR} in microblogging,'' \emph{Comput. Eng. Appl}, vol.~51, no.~8, pp. 20--26, 2015.

\bibitem{FHL+2015competing}
L.~Feng \emph{et~al.}, ``Competing for attention in social media under information overload conditions,'' \emph{PloS one}, vol.~10, no.~7, p. e0126090, 2015.

\bibitem{WLJ+2015esis}
Q.~Wang \emph{et~al.}, ``{ESIS}: {Emotion}-based spreader-ignorant-stifler model for information diffusion,'' \emph{Knowl.-Based Syst.}, vol.~81, pp. 46--55, 2015.

\bibitem{EBK+2022homophily}
G.~Ertug \emph{et~al.}, ``What does homophily do? {A} review of the consequences of homophily,'' \emph{Acad. Manag. Ann.}, vol.~16, no.~1, pp. 38--69, 2022.

\bibitem{Zmigrod2020role}
L.~Zmigrod, ``The role of cognitive rigidity in political ideologies: Theory, evidence, and future directions,'' \emph{Current Opinion in Behavioral Sciences}, vol.~34, pp. 34--39, 2020.

\bibitem{WK1994individual}
D.~M. Webster and A.~W. Kruglanski, ``Individual differences in need for cognitive closure.'' \emph{J. Pers. Soc. Psychol.}, vol.~67, no.~6, p. 1049, 1994.

\bibitem{RMR+2014information}
C.~J. Resick \emph{et~al.}, ``Information elaboration and team performance: {Examining} the psychological origins and environmental contingencies,'' \emph{Organ. Behav. Hum. Decis. Process.}, vol. 124, no.~2, pp. 165--176, 2014.

\bibitem{BAB+2018exposure}
C.~A. Bail \emph{et~al.}, ``Exposure to opposing views on social media can increase political polarization,'' \emph{Proc. Natl. Acad. Sci. U. S. A.}, vol. 115, no.~37, pp. 9216--9221, 2018.

\bibitem{WP2019elusive}
T.~Wood and E.~Porter, ``The elusive backfire effect: Mass attitudes’ steadfast factual adherence,'' \emph{Polit. Behav.}, vol.~41, pp. 135--163, 2019.

\bibitem{Hollenbaugh2021self}
E.~E. Hollenbaugh, ``Self-presentation in social media: {Review} and research opportunities,'' \emph{Rev. Commun. Res.}, vol.~9, 2021.

\end{thebibliography}
